\documentclass[12pt]{article}
\usepackage{amssymb}

\def\be{\begin{equation}}
\def\ee{\end{equation}}
\def\bea{\begin{eqnarray}}
\def\eea{\end{eqnarray}}
\def\ba{\begin{array}}
\def\ea{\end{array}}

\def\o{\omega}

\def\p{\partial}

\def\p{\partial}

\def\half{{\textstyle{\frac{1}{2}}}}

\def\np#1{{\sl Nucl.~Phys.~\bf B#1}}
\def\pl#1{{\sl Phys.~Lett.~\bf B#1}}
\def\pr#1{{\sl Phys.~Rev.~\bf D#1}}
\def\prl#1{{\sl Phys.~Rev. Lett.~\bf #1}}

\def\cqg#1{{\sl Class.~Quant.~Grav.~\bf #1}}

\catcode`\@=11
\def\@citex[#1]#2{%
\if@filesw \immediate \write \@auxout {\string \citation {#2}}\fi
\@tempcntb\m@ne \let\@h@ld\relax \def\@citea{}%
\@cite{%
  \@for \@citeb:=#2\do {%
    \@ifundefined {b@\@citeb}%
      {\@h@ld\@citea\@tempcntb\m@ne{\bf ?}%
      \@warning {Citation `\@citeb ' on page \thepage \space undefined}}%
      {\@tempcnta\@tempcntb \advance\@tempcnta\@ne%
      \@tempcntb\number\csname b@\@citeb \endcsname \relax%
      \ifnum\@tempcnta=\@tempcntb 
        \ifx\@h@ld\relax%
          \edef \@h@ld{\@citea\csname b@\@citeb\endcsname}%
        \else%
          \edef\@h@ld{\ifmmode{-}\else--\fi\csname b@\@citeb\endcsname}%
        \fi%
      \else
        \@h@ld\@citea\csname b@\@citeb \endcsname%
        \let\@h@ld\relax%
      \fi}%
    \def\@citea{,\penalty\@highpenalty\,}%
  }\@h@ld
}{#1}}

\def\@citeb#1#2{{[#1]\if@tempswa , #2\fi}}
\def\@citeu#1#2{{$^{#1}$\if@tempswa , #2\fi }}
\def\@citep#1#2{{#1\if@tempswa , #2\fi}}

\def\bcites{         
        \catcode`\@=11
        \let\@cite=\@citeb
        \catcode`\@=12
}

\def\upcites{         
        \catcode`\@=11
        \let\@cite=\@citeu
        \catcode`\@=12
}

\def\plaincites{      
        \catcode`\@=11
        \let\@cite=\@citep
        \catcode`\@=12
}

\newcount\hour
\newcount\minute
\newtoks\amorpm
\hour=\time\divide\hour by 60
\minute=\time{\multiply\hour by 60 \global\advance\minute by-\hour}
\edef\standardtime{{\ifnum\hour<12 \global\amorpm={am}%
        \else\global\amorpm={pm}\advance\hour by-12 \fi
        \ifnum\hour=0 \hour=12 \fi
        \number\hour:\ifnum\minute<10 0\fi\number\minute\the\amorpm}}
\edef\militarytime{\number\hour:\ifnum\minute<10 0\fi\number\minute}

\def\draftlabel#1{{\@bsphack\if@filesw {\let\thepage\relax
   \xdef\@gtempa{\write\@auxout{\string
      \newlabel{#1}{{\@currentlabel}{\thepage}}}}}\@gtempa
   \if@nobreak \ifvmode\nobreak\fi\fi\fi\@esphack}
        \gdef\@eqnlabel{#1}}
\def\@eqnlabel{}
\def\@vacuum{}
\def\marginnote#1{}
\def\draftmarginnote#1{\marginpar{\raggedright\scriptsize\tt#1}}
\def\draft{
        \pagestyle{plain}
        \overfullrule=2pt
        \oddsidemargin -.5truein
        \def\@oddhead{\sl \phantom{\today\quad\militarytime} \hfil
        \smash{\Large\sl DRAFT} \hfil \today\quad\militarytime}
        \let\@evenhead\@oddhead
        \let\label=\draftlabel
        \let\marginnote=\draftmarginnote
        \def\ps@empty{\let\@mkboth\@gobbletwo
        \def\@oddfoot{\hfil \smash{\Large\sl DRAFT} \hfil}
        \let\@evenfoot\@oddhead}
        \def\@eqnnum{(\theequation)\rlap{\kern\marginparsep\tt\@eqnlabel}%
        \global\let\@eqnlabel\@vacuum}  }

\title{Unitary evolution of perturbations of a two-dimensional black hole\footnote{Research supported in part by the DoE under grant DE-FG05-91ER40627.}}
\author{George Siopsis\footnote{siopsis@tennessee.edu}\\
\em Department of Physics
and Astronomy, \\
\em The University of Tennessee, Knoxville, \\
\em TN 37996 - 1200, USA.
}
\date{January 2006}
\begin{document}

\maketitle
\vspace{-3.5in}\hfill UTHET-06-0101\vspace{3.5in}

\abstract{We discuss massive scalar perturbations of a two-dimensional dilaton black hole.
We employ a Pauli-Villars regularization scheme to calculate the effect of the
scalar perturbation on the Bekenstein-Hawking entropy.
By concentrating on the dynamics of the scalar field near the horizon,
we argue that quantum effects alter the effective potential.
We calculate the two-point function explicitly and show that it
exhibits Poincar\'e recurrences.
}
\newpage

The information loss paradox following the discovery of Hawking radiation~\cite{bibH} remains unresolved despite considerable effort~\cite{bibbh1,bibbh2,bibbh3}.
In principle, quantum effects in the evolution of a black hole should be completely understood within string theory.
Since the latter is a unitary quantum theory containing gravity,
one expects that no information loss occurs during the evolution of a black
hole.
How this emerges in practice, given the existence of a horizon, remains out of
reach as calculations become intractable beyond the semi-classical approximation.

Central to our understanding of the quantum state of a black hole is the Bekenstein-Hawking entropy
which seems to be universally given by
\be\label{eqE} S_{BH} = \frac{\mathcal{A}_h}{4G} \ee
where $\mathcal{A}_h$ is the area of the horizon and $G$ is Newton's constant.
An observer outside the horizon ought to be able to understand this expression
in terms of observable matter fields.
Unfortunately, calculations generally lead to divergent expressions due to the
infinite blue-shift experienced by an in-falling object near the horizon.
To get rid of infinities, 't Hooft introduced an artificial ``brick wall'' just outside the horizon beyond which a particle cannot propagate~\cite{bibtH}.
It was subsequently understood that infinities may be absorbed by the gravitational constants
and the theory was finite when expressed in terms of physical parameters as in
any renormalizable field theory~\cite{bibE1,bibE2}.
The form~(\ref{eqE}) of the entropy, including these quantum effects, remained unchanged.

Given the finiteness of the entropy (once infinities are properly dealt with),
one expects on general grounds that the Poincar\'e recurrence theorem will hold.
If we view the matter field outside the horizon as a perturbation,
this theorem implies that, once perturbed, the system will never relax back to
its original state.
Its evolution will be quasi-periodic with a large period
\be\label{eqtP} t_P \sim O(e^{S_{BH}}) \ee
For times
$t\ll t_P$, the system may look like it is decaying back to thermal equilibrium,
but for $t\gtrsim t_P$, it should return to its original state (or close) an
infinite number of times.
This behavior should be evident in any correlator of matter fields.
In the case of an asymptotically AdS space-time,
the AdS/CFT correspondence~\cite{bibadscft} offers an additional tool in the study of unitarity,
because the CFT on the boundary of AdS is a unitary field theory~\cite{bibq13a,bibpr1,bibpr2,bibpr3,bibpr4,bibpr5,bibsol}.
It was argued by Solodukhin~\cite{bibsol} that quantum effects replace the horizon by a wormhole of narrow throat $\sim o (1/t_P)$.
It is then evident from~(\ref{eqE}) and (\ref{eqtP}) that these effects are
non-perturbative~\cite{bibego,bibsol2}.

There is no similar correspondence principle in asymptotically flat space-times.
However, the Poincar\'e recurrence theorem should still hold. To lend support
to this claim, we shall calculate the two-point function of a massive scalar
field in the background of a two-dimensional dilaton black hole.
This is a relatively simple case where explicit expressions can be derived.
By concentrating on the dynamics near the horizon, we shall argue that the
effective potential is modified by quantum effects.
For an explicit calculation, we concentrate on a wormhole modification. However,
the results are independent of the detailed shape of the effective potential.
We demonstrate that the two-point function exhibits Poincar\'e recurrences, as
expected.

The two-dimensional gravitational action is~\cite{bib2d,bibsol3}
\be S_{gr} = \frac{1}{2\pi} \int d^2 x \sqrt{-g} \ e^{-2\varphi} (R + 4(\nabla\varphi)^2 + V(\varphi)) +\frac{1}{16\pi G} \int d^2 x \sqrt{-g} \ (R-2\Lambda) \ee
The first terms provides the dynamics of the dilaton field $\varphi$.
The second term on the right-hand side is the Einstein-Hilbert action. However,
in two dimensions it does not contribute to the field (Einstein) equations,
which is why a dilaton term is needed.
For the same reason, Newton's constant $G$ is not really a coupling constant.
Nevertheless, the entropy of a two-dimensional black hole receives a contribution of the Bekenstein-Hawking form~(\ref{eqE}),
\be\label{eq4} S_{BH} = \frac{1}{4G} \ee
where the ``area'' of the horizon is $\mathcal{A}_h =1$ (single point). An additional contribution
is provided by the dilaton field $\varphi$, but will not be needed for our purposes.

A general black hole metric is
\be\label{eqmetric} ds^2 = -f(x) dt^2 + \frac{dx^2}{f(x)} \ee
The horizon is located at $x=x_h$, where
\be f(x_h) = 0 \ee
The Hawking temperature is given by
\be T_H = \frac{f'(x_h)}{4\pi} \ee
Let us add a minimally coupled massive scalar field $\phi$ of mass $m$ with action
\be\label{eqSm} S_{matter} = \frac{1}{2} \int d^2 x \sqrt{-g} \ [ (\nabla\phi)^2 - m^2 \phi^2 ] \ee
After integrating over the scalar field in the path integral, we arrive at an
effective action which is divergent.
The divergences may be eliminated by a Pauli-Villars regularization~\cite{bibE2,bibsol4}.
If we add a scalar field of (large) mass $M_1 = \sqrt{2 M^2+m^2}$ and a pair of scalar fields of mass $M_2 = \sqrt{M^2+m^2}$ each obeying
wrong statistics, we obtain the effective action
\be W_{matter} = \int d^2 x (Aa_0(x) + Ba_1(x)+\dots) \ \ , \ \ \ \
a_0 (x)=1 \ \ , \ \ \ \ a_1(x) = \frac{1}{6} R\ee
where the dots represent finite contributions and
\be A = \frac{M_1^2}{8\pi} \ln \frac{M_1^2}{M_2^2} + \frac{m^2}{8\pi} \ln
\frac{m^2}{M_2^2} \ \ ,
\ \ \ \ B = \frac{1}{8\pi} \ln \frac{M_2^4}{m^2M_1^2} \ee
are coefficients that diverge as $M\to\infty$.
They lead to a renormalization of Newton's constant $G$,
\be\label{eqGren} \frac{1}{G_R} = \frac{1}{G} + \frac{8\pi}{3} B = \frac{1}{G} + \frac{2}{3} \ln \frac{M_2^2}{mM_1} \ee
and similarly for the cosmological constant $\Lambda$.

By varying the action~(\ref{eqSm}), we obtain the wave equation for a massive scalar field in the background~(\ref{eqmetric}),
\be\label{eqscw} \frac{\p}{\p x} \left( f(x) \frac{\p\phi}{\p x} \right) - \frac{1}{f(x)}
\frac{\p^2 \phi}{\p t^2} = m^2 \phi \ee
It is convenient to introduce the ``tortoise coordinate'' $z$ given by
\be \frac{dx}{dz} = f(x) \ee
Decomposing the scalar field,
\be\label{eqscf} \phi (x,t) = \Phi (x) e^{-i\o t} \ee
the wave equation becomes
\be\label{eqwt} - \Phi'' + (m^2 f[x(z)] - \o^2) \Phi = 0 \ee 
where a prime denotes differentiation with respect to $z$.

Near the horizon, $f(x) \approx f'(x_h) (x-x_h)$, therefore
\be x-x_h \approx \frac{1}{4\pi T_H}\, e^{4\pi T_H z} \ee
where we included an arbitrary multiplicative constant for dimensional reasons. The
horizon is located at $z= - \infty$.
The wave equation near the horizon may be approximated by
\be\label{eqwh} - \Phi'' +  (m^2 e^{4\pi T_H z} - \o^2) \Phi = 0 \ee
whose solutions are Bessel functions.
Eq.~(\ref{eqwh}) is similar in form to the wave equation one obtains in higher dimensions~\cite{bibsol5}.
Demanding $\Phi \to 0$ as $z\to +\infty$, we obtain the eigenfunctions
\be \Phi_\o (z) = \mathcal{C} (\o)
\, K_{i\mu} \left( \frac{m}{2\pi T_H} \, e^{2\pi T_H z} \right)\ \ , \ \ \mu = \frac{\o}{2\pi T_H}
\ee
Notice that the spectrum is spanned by $\o \ge 0$, since $K_\nu (u) = K_{-\nu} (u)$.
To calculate the normalization factor, use
\be \int_0^\infty \frac{du}{u^{1-\epsilon}} K_{i\mu} (u) K_{i\mu'} (u) =
\frac{1}{2^{3-\epsilon} \Gamma (\epsilon)} \left| \Gamma\left( \frac{i(\mu+\mu')+\epsilon}{2} \right)
\Gamma\left( \frac{i(\mu-\mu')+\epsilon}{2} \right) \right|^2 \ee
Evidently, if $\mu \ne \mu'$, this expression vanishes as $\epsilon\to 0$ (since $\Gamma(\epsilon)\to\infty$), establishing the orthogonality of the eigenfunctions.
If $\mu'\to \mu$, we may approximate
\be \int_0^\infty \frac{du}{u^{1-\epsilon}} K_{i\mu} (u) K_{i\mu'} (u) \approx
\frac{\epsilon|\Gamma(i\mu)|^2}{2[\epsilon^2+(\mu-\mu')^2]} \ee
Taking the limit $\epsilon\to 0$, we deduce that the choice
\be \mathcal{C} (\o) = \frac{1}{\sqrt{\pi\o}\, \Gamma(-i\mu)}\ \left(\frac{m}{4\pi T_H} \right)^{-i\mu}\ee
where we included an arbitrary but convenient factor of unit norm, leads to the desirable orthogonality relation
\be \int_0^\infty dz \Phi_\o^* (z) \Phi_{\o'} (z) = \frac{1}{2\o} \delta(\o -\o') \ee
In the limit $z\to -\infty$ (as we approach the horizon), this may be approximated by plane waves,
\be\label{eqwz} \Phi_\o (z) \approx -\frac{1}{2\sqrt{\pi\o}} \left\{
e^{i\o z} - \frac{\Gamma(i\mu)}{\Gamma (-i\mu)} \left(\frac{m}{4\pi T_H} \right)^{-2i\mu} e^{-i\o z} \right\}
\ee
showing that effectively we are describing the dynamics of a massless free field
(due to its infinite blue shift) obeying the dispersion relation $\omega = |k|$.
We shall argue later that quantum effects alter this dispersion relation by an effective small mass term (corresponding to a large but finite blue shift near the horizon),
as in the case of a BTZ black hole~\cite{bibsol2}.

In the WKB approximation the wave-function for $z<z_0$, where $m^2e^{4\pi T_H z_0} - \o^2 = 0$, is
\be\label{eq24} \Phi_{\mathrm{WKB}} (z) \sim \sin \left( p(z) + \frac{\pi}{4} \right)\ \ , \ \ \ \ p(z) = \int_z^{z_0} dz' \sqrt{\o^2 -
m^2 e^{4\pi T_H z'} }\ee
We have
\be\label{eq25} p(z) = -\frac{\omega}{2\pi T_H} \left( \sqrt{1-\frac{m^2}{\omega^2} e^{4\pi T_H z}} + \ln \frac{\frac{m}{\omega} e^{2\pi T_H z}}{1+\sqrt{1-\frac{m^2}{\omega^2} e^{4\pi T_H z}}} \right)
\ee
The WKB approximation amounts to approximating a Bessel function by tangents.
In the limit $z\to -\infty$,
we may approximate
\be p(z) \approx -\omega z - \frac{\omega}{2\pi T_H} \left( \ln \frac{m}{2\omega} +1 \right) + \dots \ee
The wave-function~(\ref{eq24}) near the horizon becomes
\be\label{eqwzW} \Phi_{\mathrm{WKB}} (z) \sim e^{i\omega z} + \mathcal{S} (\omega) e^{-i\omega z} \ \ , \ \ \ \ \mathcal{S} (\omega) \approx -i \left( \frac{m}{2\omega} \right)^{-2i\mu} e^{-2i\mu}\ee
which agrees with~(\ref{eqwz}) up to an overall normalization factor,
as one can easily see by a straightforward calculation using the asymptotic Stirling expression $\Gamma (\nu) \approx \nu^{\nu -\half} e^{-\nu}$.

The free energy at temperature $T$ is
given by the WKB expression~\cite{bibtH}
\be\label{eqFdiv} F = -\frac{1}{\pi} \int_0^\infty \frac{d\o}{e^{\o /T} -1} p(-\infty) \ee
We are interested in the case where $T=T_H$, but it is convenient to work with
the above ``off-shell'' quantity in order to calculate thermodynamic quantities~\cite{bibsol4}.
The right-hand side of eq.~(\ref{eqFdiv}) is a divergent expression. However, it is not a physical
quantity; the free energy of the system has contributions from the Pauli-Villars fields as well as the gravitational field.
Adding the contributions of the Pauli-Villars regulators, we obtain a regulated expression for the free energy from matter fields
which can be written in terms of
\be p_{\mathrm{reg}} (z) = p(z) + p_{M_1}(z) - 2p_{M_2} (z)\ee
where $M_1 = \sqrt{2M^2+m^2}$, $M_2 = \sqrt{M^2+m^2}$ (recall that there is
one Pauli-Villars field of mass $M_1$ and two fields of mass $M_2$ and wrong statistics) and we defined
\be p_M(z) = -\frac{\omega}{2\pi T_H} \left( \sqrt{1-\frac{M^2}{\omega^2} e^{4\pi T_H z}} + \ln \frac{\frac{M}{\omega} e^{2\pi T_H z}}{1+\sqrt{1-\frac{M^2}{\omega^2} e^{4\pi T_H z}}} \right) \ee
so that $p_m (z) = p(z)$ (eq.~(\ref{eq25})).
In the limit $z\to -\infty$, we obtain a finite expression
\be p_{\mathrm{reg}} (-\infty) = \frac{\o}{2\pi T_H} \ln \frac{M_2^2}{mM_1} \ee
The free energy~(\ref{eqFdiv}) is corrected to
\be\label{eq32} F_{\mathrm{reg}} \equiv -\frac{1}{\pi} \int_0^\infty \frac{d\o}{e^{\o /T} -1} p_{\mathrm{reg}}(-\infty) = - \frac{T^2}{12 T_H} \ln \frac{M_2^2}{mM_1}\ee
The entropy contribution of matter fields is
\be S_{\mathrm{reg}} = - \left. \frac{\p F_{\mathrm{reg}}}{\p T} \right|_{T=T_H} =\frac{1}{6} \ln \frac{mM_1}{M_2^2} \ee
Including the gravitational contribution~(\ref{eq4}), the total entropy is
\be\label{eq34} S = S_{BH} + S_{\mathrm{reg}} = \frac{1}{4G} + \frac{1}{6} \ln \frac{mM_1}{M_2^2} = \frac{1}{4G_R} \ee
a finite quantity once expressed in terms of the physical constant $G_R$ (eq.~(\ref{eqGren})) and of
the expected Bekenstein-Hawking form.


Let us now turn to a calculation of Green functions.
We shall calculate the two-point function of the time derivative of the scalar
field, $\dot\phi$, rather than of the field $\phi$ itself, in order to avoid
unnecessary complications due to the logarithmic behavior of the two-dimensional propagator.
The two-point function at temperature $T_H$ can be written as
\be\label{eq35} G(t,z;t',z') \equiv \langle \{ \dot\phi (t,z)\,,\, \dot\phi (t',z')\} \rangle_{T_H}
= \sum_n G_0 (t+in/T_H, z; t', z')\ee
in terms of zero-temperature correlators
\be G_0 (t, z; t', z') = \int_0^\infty d\o\, \o^2\ e^{-i\o (t-t')} \Phi_\o (z) \Phi_\o^* (z') \ee
Using the approximation~(\ref{eqwz}) near the horizon (which is equivalent to the WKB expression~(\ref{eqwzW}) appropriately normalized), we obtain
\be G_0(t,z;t',z') \approx \frac{1}{2\pi (t-t'+z-z')^2} + \frac{1}{2\pi (t-t'-z+z')^2}  \ee
and after performing the sum in~(\ref{eq35}),
\be G(t,z;t',z') \approx \frac{\pi T_H^2}{2\sinh^2 \pi T_H (t-t'+z-z')} + \frac{\pi T_H^2}{2\sinh^2 \pi T_H (t-t'-z+z')}  \ee
Evidently, the two-point function decays exponentially as $t-t'\to \infty$.
This cannot be the case if the entropy of the system is finite.
To see this explicitly, let us set $z=z'$ and $t'=0$ to simplify the notation.
The two-point function behaves asymptotically for large $t>0$ as
\be\label{eq39} \mathcal{G} (t) \equiv G(t,z;0,z) \approx 4\pi T_H^2 e^{-2\pi T_H t} \ee
This expression is a valid approximation for $t \gg t_0$, where $t_0 = \frac{1}{\pi T_H}$.
The normalized time average
\be \langle |\mathcal{G}|^2 \rangle \equiv \lim_{T\to\infty} \frac{1}{T} \int_{t_0}^T dt \frac{|\mathcal{G} (t)|^2}{|\mathcal{G} (t_0)|^2} \ee
vanishes for an exponentially decaying $\mathcal{G} (t)$.
However, in a system of finite entropy $S$, one obtains
\be\label{eq41} \langle |\mathcal{G}|^2 \rangle \sim e^{-S} \ee
under general assumptions~\cite{bibGS1,bibGS2}.
In calculating the two-point function we ignored regularization issues and
therefore implicitly worked in the $S\to\infty$ limit, hence our result
$\langle |\mathcal{G}|^2 \rangle = 0$ following from eq.~(\ref{eq39}).
Quantum effects include the contributions of the Pauli-Villars fields and the gravitational field yielding a finite density of states near the horizon.
They enter the definition of Green functions and therefore alter the wave equation~(\ref{eqwh}) by modifying the effective potential.
The resulting potential must admit bound states.
To calculate these quantum effects, we shall adopt a specific form of the potential,
however the results to leading order will be independent of the details of
the shape of the effective potential.

We shall assume that the effective potential changes because of a change in the
metric due to quantum effects amounting to
\be\label{eq42} \lim_{x\to x_h} g_{tt} = -\lambda^2 \ee
thus replacing the horizon with the narrow throat of a wormhole~\cite{bibsol}.
The parameter $\lambda > 0$ is a physical parameter. We shall calculate its value
by calculating its effects on physical quantities.
On account of~(\ref{eq42})
the metric~(\ref{eqmetric}) outside the horizon changes to
\be\label{eqmetriw} ds^2 = -[f(x)+\lambda^2] dt^2 + \frac{dx^2}{f(x)} \ee
The scalar wave eq.~(\ref{eqscw}) for the field~(\ref{eqscf}) changes to
\be\label{eqscww} \sqrt{\frac{f(x)}{f(x)+\lambda^2}}\ \left( \sqrt{f(x)[f(x)+\lambda^2]}\ \Phi' \right)' + \frac{\o^2}{f(x)+\lambda^2}
\ \Phi = m^2 \Phi \ee
In terms of the ``tortoise coordinate'' $\tilde z$, where
\be \frac{dx}{d\tilde z} = \sqrt{f(x)[f(x)+\lambda^2]} \ee
we have
\be\label{eqwtw} -\Phi'' + \{ m^2 f[x(\tilde z)] + \lambda^2 m^2 - \omega^2 \} \Phi = 0 \ee
where a prime now denotes differentiation with respect to $\tilde z$,
replacing eq.~(\ref{eqwt}).

Near the horizon, $f(x) \approx 4\pi T_H (x-x_h)$. Integrating, we obtain
\be x-x_h \approx \frac{\lambda^2}{4\pi T_H}\ \sinh^2 (2\pi T_H\tilde z) \ee
and $f(x) \approx \lambda^2 \sinh^2 (2\pi T_H \tilde z)$, as long as we stay close to and outside ($x>x_h$) the horizon.
The wave equation becomes
\be -\Phi'' + \left\{ \lambda^2 m^2 \cosh^2 (2\pi T_H\tilde z) - \omega^2 \right\} \Phi = 0 \ee
If we shift
\be \tilde z \to \tilde z + \frac{1}{2\pi T_H}\, \ln \frac{2}{\lambda} \ee
the wave equation becomes
\be -\Phi'' + \left\{ m^2 e^{4\pi T_H\tilde z} + \frac{\lambda^4 m^2}{16} e^{-4\pi T_H\tilde z} + \frac{\lambda^2 m^2}{2} - \omega^2 \right\} \Phi = 0 \ee
reducing to~(\ref{eqwh}) in the limit $\lambda\to 0$.
For $\lambda >0$, the effective potential admits bound states.
In the WKB approximation, the wave-function for $|\tilde z| <\tilde z_0$, where $\lambda m \cosh (2\pi T_H\tilde z_0) = \omega$, is
\be\label{eq51} \Phi_{\mathrm{WKB}} (\tilde z) \sim \sin \left( \tilde p(\tilde z) + \frac{\pi}{4} \right) \ \ , \ \ \ \
\tilde p(\tilde z) = \int_{\tilde z}^{\tilde z_0} d\tilde z' \sqrt{\omega^2 -\lambda^2 m^2 \cosh^2(2\pi T_H \tilde z')} \ee
Eigenvalues are quantized by the Bohr-Sommerfeld quantization condition
\be\label{eq52} \tilde p (-\tilde z_0) = \left( n + \frac{1}{2} \right) \pi \ee
To integrate over $\tilde z$, change variables to $\mathbf{u}$,
\be\label{eq53} \sinh 2\pi T_H \tilde z = \frac{k}{\lambda m}\ \sin \mathbf{u} \ \ , \ \ \ \ \omega^2 = k^2 +\lambda^2 m^2\ee
We introduced the wavenumber $k$ labeling the states related to the frequency
by a dispersion relation which includes a small but finite effective mass
\be\label{eq54} m_{\mathrm{eff}} = \lambda m \ee
This reflects the fact that the new effective potential leads to large but finite blue-shifts
near the throat of the wormhole.

A short calculation yields
\bea \tilde p(\tilde z)
&=& \frac{k^2}{2\pi T_H \lambda m} \int_{\mathbf{u}}^{\pi /2}
\frac{\cos^2 \mathbf{u}' d\mathbf{u}'}{\sqrt{ 1+ \frac{k^2}{\lambda^2 m^2} \sin^2 \mathbf{u}'}} \nonumber\\
&=& \frac{\o}{2\pi T_H}\ \left\{ \mathbf{K} (\frac{k}{\o} )-\mathbf{E} (\frac{k}{\o} ) -F(\alpha, \frac{k}{\o} ) + E(\alpha, \frac{k}{\o} ) - \frac{k^2}{\o^2} \sin\alpha \cos \mathbf{u} \right\} \eea
written in terms of elliptic functions, where
\be\sin\alpha = \frac{\o}{\lambda m}\, \frac{\sin\mathbf{u}}{\sqrt{1+ \frac{k^2}{\lambda^2 m^2} \sin^2 \mathbf{u}}} = \frac{\o}{k} \tanh 2\pi T_H \tilde z\ee
For $k\approx \o$, we may expand
\be\label{eqFE} F(\alpha , \frac{k}{\o} ) - E(\alpha , \frac{k}{\o} ) = \frac{2}{\pi} \mathbf{E}' (\frac{k}{\o} )
\ln\tan \left( \frac{\alpha}{2} + \frac{\pi}{4} \right) + O(\alpha)\ee
and also
\be\label{eqKE} \mathbf{K} (\frac{k}{\o} ) - \mathbf{E} (\frac{k}{\o} ) = \ln \frac{4\o}{\lambda m} -1 +\dots \ee
In the WKB approximation, the free energy at temperature $T$ is
\be F = - \frac{1}{\pi} \int_0^\infty \frac{d\omega}{e^{\omega /T} - 1}\ \tilde p (-\tilde z_0) = - \frac{1}{\pi^2 T_H} \int_0^\infty \frac{d\omega\, \o}{e^{\omega /T} - 1}\ \left\{ \mathbf{K} (\frac{k}{\o} )-\mathbf{E} (\frac{k}{\o} )
\right\} \ee
Using the approximation~(\ref{eqKE}), after some algebra we obtain
\be F = - \frac{T^2}{12T_H} \left\{ \ln\frac{1}{\lambda} +\psi(2)+\ln \frac{4T}{m}
 -1 +\frac{12\zeta'(2)}{\pi^2} \right\}
\ee
a finite expression, replacing the infinite expression~(\ref{eqFdiv}).
However, it should be emphasized that $\lambda$ is not a regularization parameter but a physical one and the above expression is not a physical quantity
by itself.
We ought to add the contribution of the Pauli-Villars fields in order to obtain the contribution of matter fields to the free energy.
We readily obtain
\be F_{\mathrm{reg}} = - \frac{T^2}{12T_H} \ln \frac{M_2^2}{mM_1} \ee
in agreement with our earlier result~(\ref{eq32}).
Together with the gravitational contribution, they form a physical quantity
from which one can deduce thermodynamical quantities such as the entropy~(\ref{eq34}) of the system.
Notice that the dependence on $\lambda$ has disappeared so no conclusion on its value can
be drawn by calculating the free energy of the system and consequently its entropy.

To calculate $\lambda$, we turn to a calculation of the two-point function.
In the WKB approximation, Green functions may be expressed in terms of the
wave-functions~(\ref{eq51}).
Using~(\ref{eqFE}) and (\ref{eqKE}), we obtain after some algebra
\be \Phi_{\mathrm{WKB}} (\tilde z) \sim e^{ik\tilde z} + \tilde\mathcal{S} (\omega) e^{-ik\tilde z} \ \ , \ \ \ \ |\tilde\mathcal{S} (\omega)|^2 = 1 \ee
which is similar to our earlier result~(\ref{eqwzW}) with the crucial difference that the dispersion relation has changed from a massless one ($\o = |k|$) to eq.~(\ref{eq53}) which includes a small effective mass (eq.~(\ref{eq54})).
The explicit form of $\tilde\mathcal{S} (\o)$ is not needed; only the fact that it is of unit norm.
On account of~(\ref{eqKE}), the quantization condition~(\ref{eq52}) reads
\be \frac{k}{\pi T_H} \left( \ln \frac{4k}{\lambda m} - 1 +\dots \right) =
\left( n_k + \frac{1}{2} \right) \pi \ \ , \ \ \ \ n_k\in\mathbb{Z}\ee
therefore
\be k \approx \left( n + \frac{1}{2} \right) \frac{\pi^2 T_H}{\ln\frac{1}{\lambda}} \ee
It follows that the two-point function is periodic under
\be\label{eq65} \tilde z \to \tilde z + \tilde n L_{\mathrm{eff}} \ \ , \ \ \ \ L_{\mathrm{eff}} = \frac{2\ln\frac{1}{\lambda}}{\pi T_H} \ \ , \ \ \ \
\tilde n \in \mathbb{Z} \ee
Ignoring temperature effects, we may write an expression for it using the method of images,
\be G_0 (t,z;t',z') = \sum_{\tilde n} \int_0^\infty dk\, k\, e^{-i\o (t-t')} \Phi_{\mathrm{WKB}} (\tilde z+ \tilde n L_{\mathrm{eff}}) \Phi_{\mathrm{WKB}}^*(\tilde z') \ee
We shall calculate this correlator following a similar calculation in~\cite{bibsol2}.
Each term in the series can be written in terms of the two-point function of
a massive field of mass $m_{\mathrm{eff}} = \lambda m$ on account of the dispersion relation~(\ref{eq53}).
Setting $z'=z$ and $t'=0$ for simplicity, we obtain
\be \mathcal{G}_0 (t) \equiv G_0 (t,z; 0,z) = -\frac{1}{2} \sum_{\tilde n} \ddot H_0^{(2)} (\lambda m\sqrt{t^2 + \tilde n^2 L_{\mathrm{eff}}^2}) \ee
For small $t\lesssim t_0 = \frac{1}{\pi T_H}$, only the $\tilde n = 0$ contributes (in the other terms, the argument of the Hankel function is approximately
constant, so the time derivative vanishes).
We obtain
\be\label{eq68} \mathcal{G}_0 (t) \approx \frac{1}{\pi t^2} \ee
exhibiting a power law decaying behavior for large $t$, which turns into an
exponential decay (eq.~(\ref{eq39})) once temperature effects are included.

For large $t$ ($t\gtrsim t_0$), however, this approximation is no longer valid. For $t\gg \frac{1}{\lambda m}$, we
may approximate the sum by an integral.
After some algebra, we arrive at~\cite{bibsol2}
\be\label{eq69} \mathcal{G}_0 (t) \approx \frac{\pi\lambda m}{2L_{\mathrm{eff}}} e^{-i\lambda mt} \ee
exhibiting periodicity with period (Poincar\'e time)
\be\label{eq70} t_P = \frac{2\pi}{\lambda m}\ee
Including temperature effects does not alter the above result of periodicity
because the Green function at finite temperature may be written as a series (eq.~(\ref{eq35})) each term of which is periodic with period given by eq.~(\ref{eq70}).
The normalized time average is
\be\label{eq71} \langle |\mathcal{G}_0|^2 \rangle \equiv \lim_{T\to\infty} \frac{1}{T} \int_{t_0}^T dt \frac{|\mathcal{G}_0 (t)|^2}{|\mathcal{G}_0 (t_0)|^2}  \sim \frac{\lambda^2}{\ln^2 \frac{1}{\lambda}} \ee
where in the last step we used eqs.~(\ref{eq68}) and (\ref{eq69}) together with the definition~(\ref{eq65}) of $L_{\mathrm{eff}}$.
Comparing with the general result~(\ref{eq41})~\cite{bibGS1,bibGS2}, it follows that asymptotically,
\be \lambda^2 \approx e^{-S} = e^{- \frac{1}{4G_R}} \ee
where we used~(\ref{eq34}). The Poincar\'e time~(\ref{eq70}) to leading order is given by
\be\label{eq73} t_P \approx \frac{2\pi}{m}\ e^{\frac{1}{8G_R}}\ee
In conclusion, we calculated the two-point function of a two-dimensional massive scalar field
in a black hole background.
We argued that quantum effects altered the effective potential replacing the
horizon by the narrow throat of a wormhole,
following a similar argument in three-dimensional asymptotically AdS space by
Solodukhin~\cite{bibsol}.
In the latter case, the throat size could be deduced by going to the boundary of AdS and applying the AdS/CFT correspondence~\cite{bibego,bibsol2}.
We were able to find an expression for the throat of the wormhole in our case
by concentrating on the dynamics near the horizon (throat).

Although we explicitly considered an effective potential arising from a
narrow wormhole,
our results are independent of the details of the potential.
To leading order, the oscillatory behavior~(\ref{eq69}) relies solely on the
dispersion relation~(\ref{eq53}) in which the effective mass~(\ref{eq54}) is determined
by the minimum of the effective potential.
Moreover, the
effective length~(\ref{eq65}) which determines the spatial periodicity of Green functions and enters the normalized average~(\ref{eq71})
is also generically given to leading order by $L_{\mathrm{eff}} \sim \ln \frac{1}{\lambda}$, therefore the expression~(\ref{eq73}) for the Poincar\'e time holds more generally.
These observations support the argument that by modifying the effective potential
we have correctly accounted for quantum effects to leading order.

It would be interesting to generalize the discussion to higher-dimensions.
We should obtain qualitatively similar results because the potential near the
horizon retains the same form~\cite{bibsol5}.
By concentrating on the dynamics near the horizon, we may thus arrive at a {\em quantitative} understanding of Poincar\'e recurrences
in an asymptotically flat black hole background.

%


\end{document}